\documentclass[aps,prb,reprint]{revtex4-1}

\usepackage{amsmath,amsfonts,amssymb,graphicx,braket}
\usepackage[]{units}
\usepackage{hyperref}
\hypersetup{
    colorlinks=true,
    linkcolor=blue,
    filecolor=blue,      
    urlcolor=blue,
    citecolor=blue
}

\begin{document}
\title{Spin-Degenerate Regimes for Single Quantum Dots in Transition Metal Dichalcogenide Monolayers}
\author{Matthew Brooks}
\email{matthew.brooks@uni-konstanz.de}
\author{Guido Burkard}
\affiliation{Department of Physics, University of Konstanz, D-78464, Germany}
%\date{\today}

\begin{abstract}
Strong spin-orbit coupling in transition metal dichalcogenide (TMDC) monolayers results in spin resolvable band structures about the $K$ and $K'$ valleys such that the eigenbasis of a 2D quantum dot (QD) in a TMDC monolayer in zero field is described by the Kramers pairs $\ket{0}_-=\ket{K'\uparrow}$, $\ket{1}_-=\ket{K\downarrow}$ and $\ket{0}_+=\ket{K\uparrow}$, $\ket{1}_+=\ket{K'\downarrow}$. The strong spin-orbit coupling limits the usefulness of single TMDC QDs as spin qubits. Possible regimes of spin-degenerate states, overcoming the spin-orbit coupling in monolayer TMDC QDs are investigated in both zero field, where the spin and valley degrees of freedom become fourfold degenerate, and twofold degeneracy in some magnetic field, localised to a given valley. Such regimes are shown to be achievable in MoS$_{2}$, where the spin orbit coupling is sufficiently low and of the right sign such that the spin resolved conduction bands intersect at points about the $K$ and $K'$ valleys and as such may be exploited by selecting suitable critical dot radii. 
\end{abstract}
\maketitle

\section{Introduction}
 \label{sec:Intro}
 
\begin{figure}[!h]
    \includegraphics[width=\linewidth]{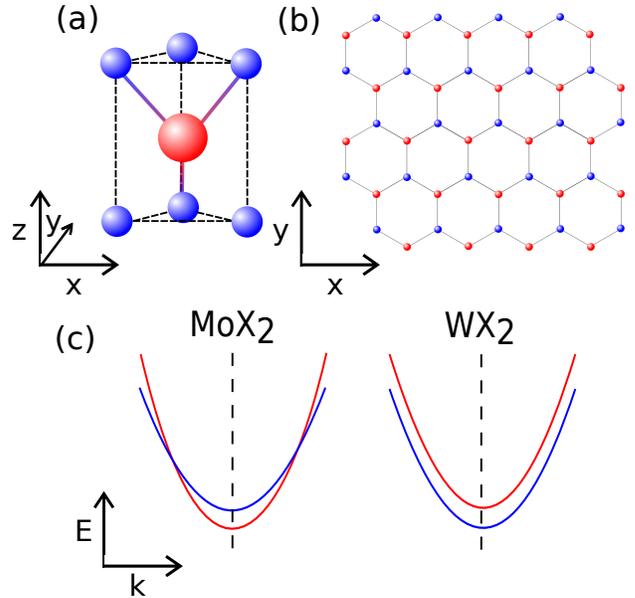}
    \caption{(a) 3D view of a TMDC unit cell (red denoting M atoms, blue denoting X atoms) showing the three sub layers of a TMDC monolayer and the broken inversion of the crystal lattice. (b) Planar (X-Y) view of a TMDC lattice. (c) Spin resolved conduction band (red: $\ket{0}_-=\ket{K'\uparrow}$ and $\ket{1}_-=\ket{K\downarrow}$, blue: $\ket{0}_+=\ket{K\uparrow}$ and $\ket{1}_+=\ket{K'\downarrow}$) around the $K$ valley in the BZ of Mo and W based TMDC monolayers demonstrating the spin crossings present in Mo TMDCs and not in W, the $K'$ valley may be visualised simply by the time-reversal of the given band structure.}
    \label{fig:TMDC_Cartoon}
\end{figure}

Transition metal dichalcogenide (TMDC) monolayers are atomically thin crystal layers exfoliated down from bulk weakly cohesive stacks. Similarly to graphene, a hexagonal lattice of alternating lattice sites results in two inequivalent, time-reversal symmetric valleys ($K$ and $K'$), see Fig. \ref{fig:TMDC_Cartoon}~(b) \cite{wang2012electronics,suzuki2014valley,cao2011mos_2,kormanyos2015k}. Unlike graphene, the monolayer crystals posses broken inversion symmetry, see Fig. \ref{fig:TMDC_Cartoon}~(a), inducing direct band gaps in the visible range about the two valleys \cite{splendiani2010emerging,mak2010atomically,lu2013intervalley}. Furthermore, strong spin-orbit coupling from the transition metal atoms introduces a strong coupling between the spin and valley degrees of freedom, see Fig. \ref{fig:TMDC_Cartoon}~(a) \cite{xiao2012coupled,shan2013spin,xu2014spin}. TMDCs are characterised by the chemical composition MX$_{2}$, where M denotes the transition metal (Mo or W) and X denotes the chalcogenide (S or Se). The presence of a direct band gap and spin-valley coupling in a two-dimensional material allows for a number of interesting electronic, spintronic and valleytronic applications including room temperature quantum spin Hall insulators, optically pumped valley polarisation, long lived exciton spin polarisation and 2D quantum dots (QDs) \cite{yang2015long,PhysRevB.93.035442,cazalilla2014quantum,PhysRevX.4.011034,PhysRevB.93.045313,zeng2012valley}. 

While the strong spin-valley coupling of TMDC monolayers offers numerous interesting physical phenomena, it presents a difficulty for qubit implementation in gated QDs. Kramers pairs of the spin and valley degrees of freedom result from this coupling \cite{cazalilla2014quantum,PhysRevX.4.011034,klinovaja2013spintronics}. At low energy the $\ket{0}_-=\ket{K'\uparrow}$ and $\ket{1}_-=\ket{K\downarrow}$ states are degenerate in zero field and are energetically separate from the $\ket{0}_+=\ket{K\uparrow}$ and $\ket{1}_+=\ket{K'\downarrow}$ states \cite{PhysRevX.4.011034,song2013transport,kormanyos2013monolayer}. This effect can be observed in the spin resolvable structure of the conduction band (CB) about the $K$($K'$) points \cite{kosmider2013large,zhu2011giant} as shown in Fig. \ref{fig:TMDC_Cartoon}~(c). The obvious choice for the computational basis of a qubit is therefore a spin-valley qubit consisting of the two states of the lowest lying Kramers pair, $\ket{0}_-(\ket{1}_-)$ in Mo\textit{X}$_{2}$ and $\ket{0}_+(\ket{1}_+)$ in W\textit{X}$_{2}$, where the required energy difference may be achieved by spin-valley Zeeman splitting induced by a perpendicular magnetic field \cite{flensberg2010bends,aivazian2015magnetic,srivastava2015valley,rostami2015valley}. However, such qubits are inherently limited by a necessity for coupling of the valley states. Methods of doing so have been proposed in carbon nanotubes by means of short range disorder in the dots \cite{flensberg2010bends,palyi2011disorder}, requiring atomic level engineering, or by optical manipulation\cite{ye2016optical}. Additionally, the valley coherence of WSe$_{2}$ excitons has been measured\cite{hao2016direct}, demonstrating an order of magnitude lower coherence times than spin in other TMDC monolayer crystals\cite{yang2015long}. If qubits in TMDC monolayers could operate similarly to semiconductor spin qubits then the broad theoretical and experimental findings of the field \cite{hanson2007spins,zwanenburg2013silicon,petta2005coherent} may be directly utilised. In so doing, a novel breed of 2D, optically active, direct band gap, and relatively nuclear spin free\cite{PhysRevB.93.045313} semiconductor spin qubits are gained without the need for an artificially induced band gap, as is needed in graphene\cite{zhou2007substrate}. This requires a method of manipulating the dots such that the spin-orbit coupling may be suppressed and regimes of pure spin qubits may be accessed.  

There is a noticeable and useful difference between the low energy band structures of Mo based and W based monolayers as demonstrated in Fig. \ref{fig:TMDC_Cartoon}~(c): the band crossings observed in the spin resolved CB structures in Mo monolayers which are absent in W monolayers which suggest that it is possible to achieve spin degeneracy localised within a given valley. Such spin-degenerate regimes offer the possibility of implementing the desired pure spin qubits in TMDCs. Additionally, by placing a TMDC material in a perpendicular magnetic field, breaking time reversal symmetry, valley Zeeman splitting may be introduced to the system. Previous work\cite{PhysRevX.4.011034} has suggested that it may be possible to access regimes of spin degeneracy within the same valley by introducing a large magnetic field. In this work, we build upon previous analyses of  TMDC QDs in a effective low energy regime by solving for various conditions in which a spin qubit may be viable, demonstrating a dot size tuneable spin-orbit splitting and investigating the effects of a finite potential well model as opposed to previous assumptions of an infinite potential.

Here, we present methods of achieving spin degeneracy within a given valley of a QD in a TMDC monolayer at zero or moderate fields. Firstly in Sec. \ref{sec:ZF} a zero external field model is discussed, demonstrating the Kramers pairing of states as to derive an expression for a critical radius at which fourfold spin-valley degeneracy may be expected. Also we discuss the best candidate monolayer for a pure spin qubit. Then in Sec. \ref{sec:PerpB} an external magnetic field perpendicular to the dot is considered and numerical solutions to the necessary external field strengths at a given dot radius are shown at which a spin-degenerate state within a given valley is expected. Next the effects of finite confinement potential are shown on the two previously discussed regimes is given in Sec. \ref{sec:FinWell}. Finally, an effective implementation regime for the various methods of achieving valley independent spin degeneracy is discussed in Sec. \ref{sec:QIP} before a summary is given in \ref{sec:Summary}.

\section{Zero Field}
 \label{sec:ZF}

To describe a QD in monolayer TMDC the following effective low energy Hamiltonian about the $K$ and $K'$ point in the CB is employed\cite{PhysRevX.4.011034}

\begin{equation}
    H_{\text{dot}}=H_{\text{el}}^{\tau,s}+H_{\text{so}}^{\text{intr}}+V=\frac{\hbar^2q_+q_-}{2m_{\text{eff}}^{\tau,s}}+\tau\Delta_{\text{cb}}s_z+V.
    \label{eq:Total_Hamiltonian}    
\end{equation}

\noindent Here, $\tau=1(-1)$ refers to the $K$ and $K'$ valley, $s_z$ gives the spin Pauli-z matrix with eigenvalues $s=1(-1)$ for spin $\uparrow(\downarrow)$, wave number operators $q_\pm=q_x\pm iq_y$ where $q_k=-i\partial_k$, $\Delta_{cb}$ is the energy spliting in the CB due to the strong intrinsic spin-orbit coupling of the TMDC monolayer and the spin-valley dependant effective electron mass is defined as $1/m_{\text{eff}}^{\tau,s}=1/m_{el}^0-\tau s/\delta m_{\text{eff}}$ where $\delta m_{\text{eff}}$ is material dependant. Initially, it is assumed that the QD potential $V$ is sufficiently deep such that it may be described by an infinite hard walled potential 

\begin{equation}
V=\begin{cases}
    0       & \quad r\leq R_D\\
    \infty  & \quad r>R_D\\
  \end{cases}
  \label{eq:Infin_Pot}
\end{equation}

\noindent where $r$ is the radial coordinate and $R_D$ is the radius of the dot. This may be assumed  in lieu of a harmonic potential, as is often used in bulk semiconductor QD models, since the 2D nature of a TMDC allows for a more direct interface between the gates and the plane in which an electron will be confined. Additionally, such an assumption allows for edge effects at the boundary of the dot to be neglected. In 2D polar coordinates, the wave number operators may be defined as 

\begin{equation}
    q_\pm=\pm i e^{\pm i\phi}(\mp\partial_r-\frac{i}{r}\partial_{\phi}).
    \label{eq:radial_Wavenumber}
\end{equation}

\noindent where $\phi$ is the angular coordinate. Assuming the dot to be circular, rotational symmetry about the z-axis dictates that the dot's Hamiltonian will commute and share eigenstates with the z-component of the angular momentum operator ($l_z$). This allows for the normalised solution of the angular component of the wavefunction $\Psi(r,\phi)=R(r)\Phi(\phi)$ to be given as

\begin{equation}
    \Phi(\phi)=\frac{e^{il\phi}}{\sqrt{2\pi}}.
    \label{eq:angular_Comp_WVFN}
\end{equation}

\noindent Since the radial component of the wavefunction observes the boundary condition $R(R_D)=0$, the following expression is derived where $j_{n,l}$ is the $n^{th}$ zero ($n=1,2,3,\dots$) of the $l^{th}$ Bessel function of the first kind $J_{l}$ ($l=0,\pm1,\pm2,\dots$)  

\begin{equation}
    R_{n,l}(r)= \frac{(-1)^ {\frac{|l|-l}{2}}\sqrt{2}J_{|l|}\left(\frac{j_{n,|l|}}{R_D}r\right)}{R_Dj_{n,|l|+1}}.
    \label{eq:radial_ZF_WVFN}   
\end{equation}

\noindent As such, the full normalised solutions of a hard wall TMDC quantum dot in zero external field are given in the spinor form as

\begin{subequations}
    \begin{align}
        \Psi^{\uparrow}_{n,l}(r,\phi)&=\frac{e^{il\phi}}{\sqrt{2\pi}}\left(\begin{array}{c}1\\0\\\end{array}\right)R_{n,l}(r),\\
        \Psi^{\downarrow}_{n,l}(r,\phi)&=\frac{e^{il\phi}}{\sqrt{2\pi}}\left(\begin{array}{c}0\\1\\\end{array}\right)R_{n,l}(r),
    \end{align}
    \label{eq:Total_Spinors}
\end{subequations}

\noindent and the spin, valley and dot radius dependant energy eigenvalues are given as
 
\begin{equation}
    E^{n,l}_{\tau,s}(R_D)=\frac{\hbar^2j_{n,|l|}^2}{2m^{\tau,s}_{\text{eff}}R_D^2}+\tau s \Delta_{\text{cb}}.
    \label{eq:ZF_Energy}
\end{equation}

\begin{figure}[!h]
    \includegraphics[width=\linewidth]{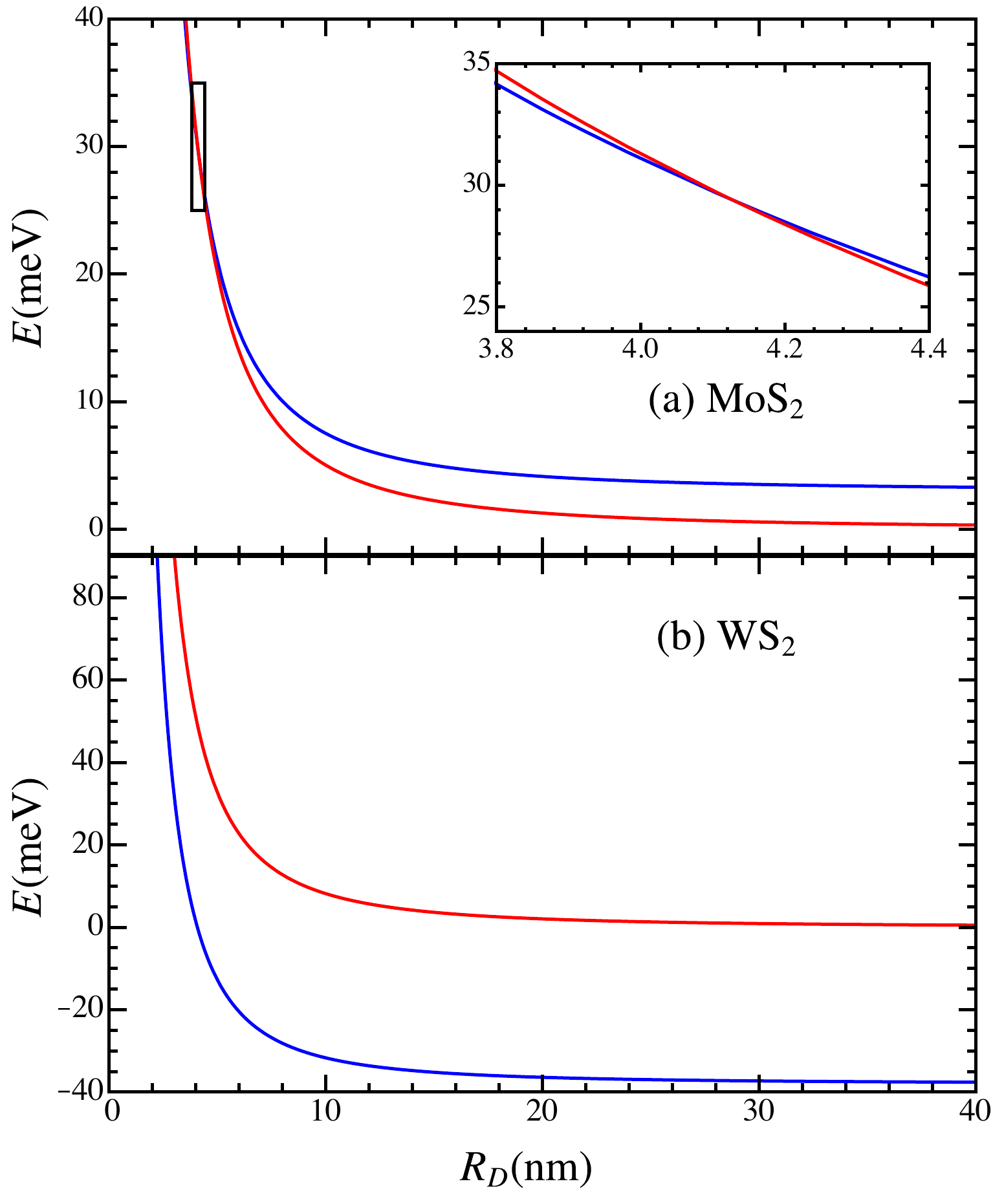}
    \caption{(a) Zero field energy spectrum of the $n=1$, $l=0$ eigenstates, blue: $\ket{0}_+(\ket{1}_+)$ and red: $\ket{0}_-(\ket{1}_-)$, of MoS$_{2}$ hard wall QD of a given dot radius $R_D$, here a point of fourfold degeneracy of the valley-spin eigenstates is observed at a particular radius. Inset: region about which the fourfold degeneracy is observed in the spectrum. (b) Zero field energy spectrum of the $n=1$, $l=0$ eigenstates of WS$_{2}$ hard wall QD of a given dot radius, here no point of fourfold degeneracy of the valley-spin eigenstates is observable due to the $\Delta_{cb}>0$ not being satisfied by W based TMDCs.}
    \label{fig:MoS2_ZF_GS}
\end{figure}

From the four realisations of spin and valley, only two separate energy solutions in zero field emerge, i.e. $E^{n,l}_{K,\uparrow}=E^{n,l}_{K',\downarrow}=E^{n,l}_{+}$ and $E^{n,l}_{K',\uparrow}=E^{n,l}_{K,\downarrow}=E^{n,l}_{-}$. These two possible solutions describe the  $\ket{0}_+(\ket{1}_+)$ and $\ket{0}_-(\ket{1}_-)$ Kramers pairs respectively. If the two solutions are assumed to be equivalent, then Eq. (\ref{eq:ZF_Energy}) may be used to describe the radius at which fourfold degeneracy in the valley-spin Hilbert space is achieved. As such, a critical radius $R^{n,l}_{c}$ at which $E^{n,l}_{+}=E^{n,l}_{-}$ is given by

\begin{equation}
    R^{n,l}_{\text{c}}=\frac{\hbar j_{n,|l|}}{2\sqrt{\Delta_{\text{cb}}}}\sqrt{\frac{1}{m^{-}_{\text{eff}}}-\frac{1}{m^{+}_{\text{eff}}}}
    \label{eq:ZF_Crit_Rad}
\end{equation}

\noindent where $m^{-}_{\text{eff}}=m^{K\downarrow/K'\uparrow}_{\text{eff}}$ and $m^{+}_{\text{eff}}=m^{K\uparrow/K'\downarrow}_{\text{eff}}$. Therefore, there are real solutions to the critical radius at which fourfold valley-spin degeneracy may exist for dots with intrinsic spin-orbit coupling such that $\Delta_{cb}>0$ and $m^{+}_{\text{eff}}>m^{-}_{\text{eff}}$. The latter condition is given for all possible TMDC monolayers while the former is only satisfied by Mo based TMDCs ($\Delta_{cb}=\unit[1.5]{meV}$  for MoS$_{2}$ and $\Delta_{cb}=\unit[11.5]{meV}$ for MoSe$_{2}$) \cite{kormanyos2015k,PhysRevX.4.011034} (see Fig. \ref{fig:MoS2_ZF_GS}). Alternatively, real solutions of $R_{\text{c}}$ may be found in materials where both $\Delta_{cb}<0$ and $m^{+}_{\text{eff}}<m^{-}_{\text{eff}}$, however, there is no known TMDC that satisfies the latter condition.

In the groundstate ($n=1$, $l=0$) the critical radius at which fourfold degeneracy may be expected is $\unit[4.13]{nm}$ for MoS$_{2}$ and $\unit[1.46]{nm}$ for MoSe$_{2}$ QDs. While both radii are difficult to achieve by electrostatic gating, MoS$_{2}$  monolayers offer plausibly achievable fourfold degeneracy through some critical radii and consequently prove themselves as a the most viable candidate for 2D single QD pure spin qubits. For the remainder of the presented work we will focus solely on MoS$_{2}$ monolayers.

\section{Perpendicular Magnetic Field}
 \label{sec:PerpB}
 
Following the previous methods\cite{PhysRevX.4.011034}, the spin-valley eigenenergies of a TMDC monolayer QD in a constant perpendicular magnetic field ($B_z$) may be derived from the following Hamiltonian
\begin{equation}
\begin{split}
   H_{B_\perp}^{\tau,s}= \hbar\omega_c^{\tau,s}\alpha_+\alpha_-+\tau\Delta_{\text{cb}}s_z+\frac{1+\tau}{2}\frac{B_z}{|B_z|}\hbar\omega_c^{\tau,s}   \\ 
    +\frac{1}{2}(\tau g_{\text{vl}}+g_{\text{sp}} s_z )\mu_B B_z& 
    \end{split}
    \label{eq:BPerp_Hamiltonian}
\end{equation}  
where the cyclotron frequency is defined as $\omega_c^{\tau,s}=e|B_z|/m_{\text{eff}}^{\tau,s}$, $\mu_B$ is the Bohr magneton, $g_{sp}$ is the spin g-factor, $g_{vl}$ is the valley g-factor and $\alpha_{\pm}$ denote the modified wavenumber operators $\alpha_{\pm}=\mp i l_Bq_{\pm}/\sqrt{2}$ where $l_B=\sqrt{\hbar/eB_z}$ is the magnetic length. After appropriate gauge selection wavefunctions in terms of the dimensionless length parameter $\rho=r^2/2l_B^2$ are given as $P_{n,l}(\rho)= \rho^{|l|/2} e^{-\rho/2}M (a_{n,l},|l|+1,\rho)$ where $a_{n,l}$ describes the $n^{th}$ solution of the following bound state identity $M(a_{n,l},|l|+1,\rho_D)=0$, where $\rho_D=\rho[r=R_D]$ and $M(a,b,c)$ is the confluent hypergeometric function of the first kind. The addition of an out of plane magnetic field does not break the rotational symmetry of the dot, hence the angular component of the wavefuntion is not affected by this change. The eigenenergies are therefore given as

\begin{equation}
    \begin{split}
       E^{\tau,s}_{n,l}= \hbar\omega_c^{\tau,s}\left(\frac{1+\tau}{2}\frac{B_z}{|B_z|}+\frac{|l|+l}{2}-a_{n,l}\right) \\ +\tau\Delta_{\text{cb}}s_z+\frac{1}{2}(\tau g_{\text{vl}}+sg_{\text{sp}} )\mu_BB_z.  
        \end{split}
        \label{eq:BPerp_total_Energy}
\end{equation}

\begin{figure}[!h]
    \includegraphics[width=\linewidth]{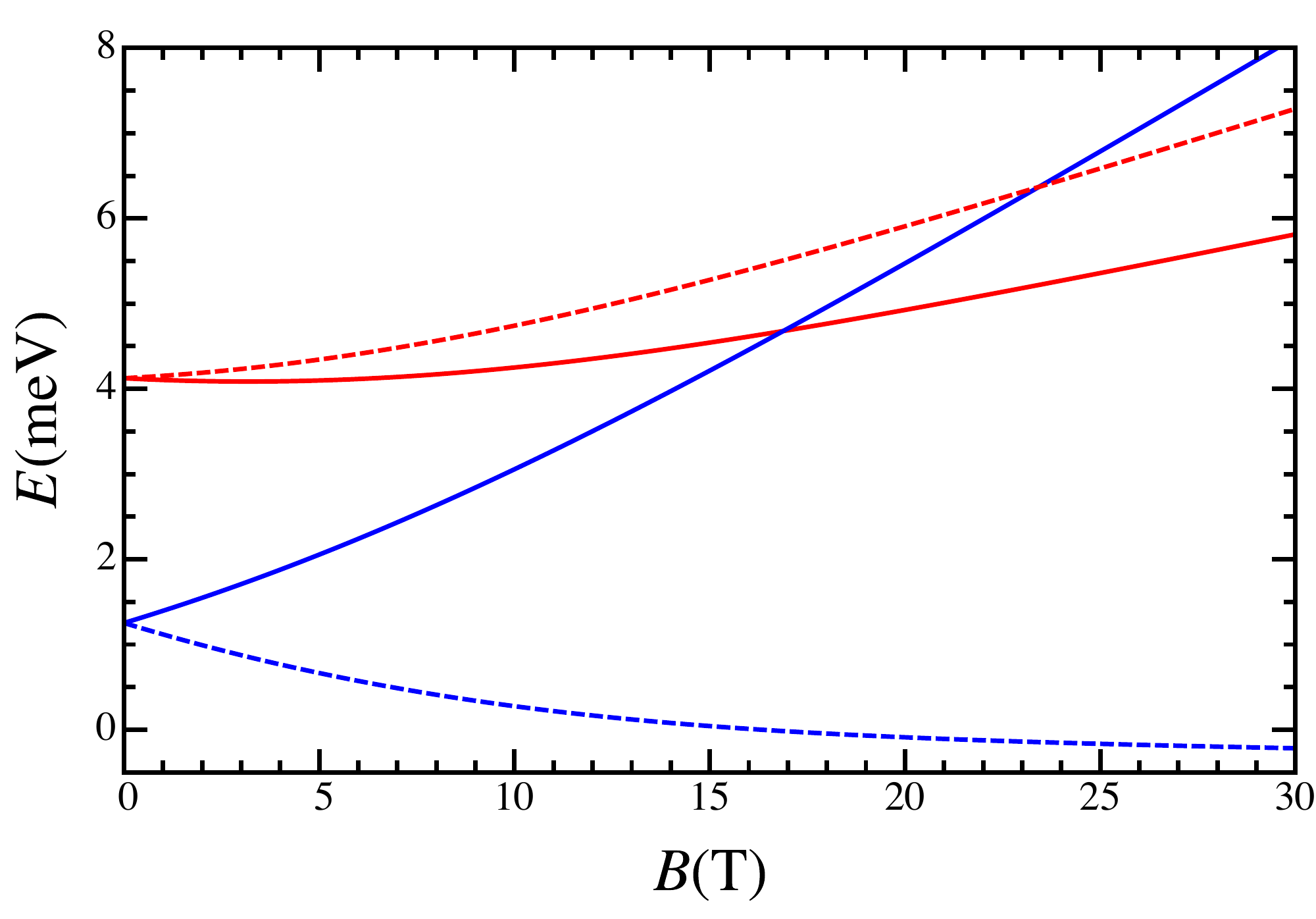}
    \caption{Energy spectra of the $n=1$, $l=0$ state in a QD of $\unit[20]{nm}$ radius on a MoS$_{2}$ monolayer with under a perpendicular magnetic field. Here the critical field strength at which $E^{n=1,l=0}_{K',\downarrow}=E^{n=1,l=0}_{K',\uparrow}$ is observed at the high magnetic field strength of $\sim\unit[23]{T}$. Blue solid (dashed) line: $\ket{K'\uparrow}$ ($\ket{K\downarrow}$) and red solid (dashed) line: $\ket{K\uparrow}$ ($\ket{K'\downarrow}$).}
    \label{fig:MoS2_Inf_BPerp_20nm}
\end{figure}

\begin{figure}[!h]
    \includegraphics[width=\linewidth]{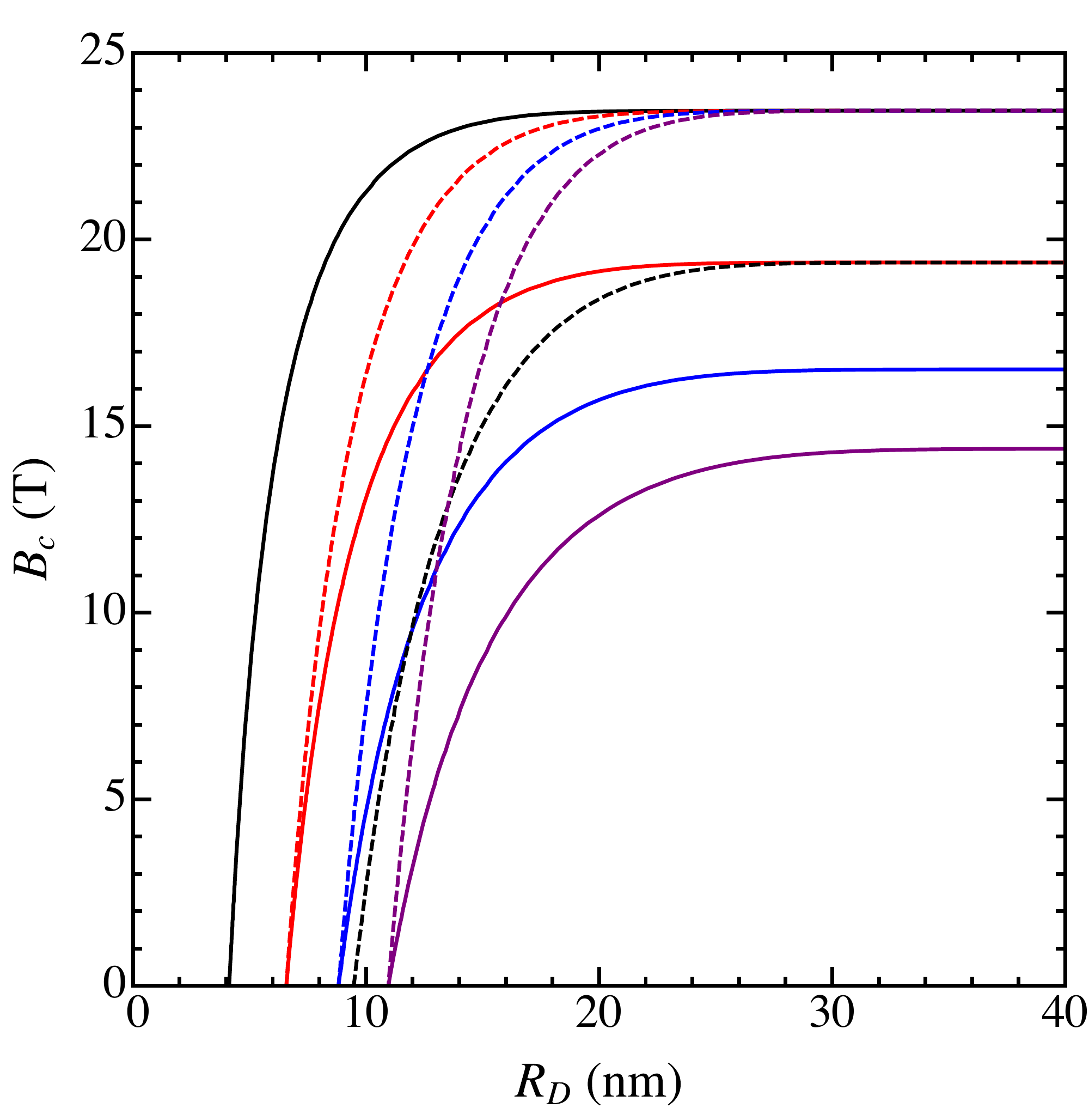}
    \caption{Spin degeneracy curves of critical out of plane magnetic field strength $B_c$ with the radius of QD on MoS$_{2}$ monolayer for the first few states, black solid (dashed): $n=1$ ($2$), $l=0$, red solid (dashed): $n=1$, $l=1$ ($-1$), blue solid (dashed): $n=1$, $l=2$ ($-2$), purple solid (dashed): $n=1$, $l=2$ ($-2$).}
    \label{fig:MoS2_Inf_BCrit_RDot_Spin}
\end{figure}

From Eq. (\ref{eq:BPerp_total_Energy}), spectra demonstrating the effect of an out of plane magnetic field for QDs in MoS$_{2}$ monolayers may be calculated numerically. The splitting of the spin and valley states due to the external magnetic field allows for spin-degenerate crossings for a given radius within the $K'$ valley, i.e. at some external magnetic field strength $E^{n,l}_{K',\uparrow}=E^{n,l}_{K',\downarrow}$, see Fig. \ref{fig:MoS2_Inf_BPerp_20nm}. These critical magnetic field strengths ($B_{\text{c}}$) for given dot radii may be determined for a range of radii to give the spin-degenerate regime spectra shown in Fig. \ref{fig:MoS2_Inf_BCrit_RDot_Spin}.

These spectra show separate plateaus in the critical field strength at relatively high dot radii ($R>\unit[20]{nm}$) for the $l\geq0$ and $l<0$ angular states, differing by up to $\sim\unit[5]{T}$, but with both still at high field strengths. This is the limit at which the maximum Kramers pair energy difference at zero field is observed and valley Zeeman splitting alone is used to achieve spin degeneracy. On the other end of the spectra, at low external field strengths the gradient of the regime curves increases compromising the fabrication error robustness of single dot spin qubits, i.e. small errors ($\sim\unit[1]{nm}$) in QD radii would make the difference between operating the qubit at $\unit[1]{T}$ and $\unit[6]{T}$ external field. Thus operating a spin qubit with a single electron regime in the groundstate is not easily implemented. The possibility of operation at excited states and alternative enhancment methods are considered and discussed in Sec. \ref{sec:QIP}.

\section{Finite Well}
 \label{sec:FinWell}
 
 Up to this point, all models used assume QDs with an infinite hard wall potential. Here the effects of transitioning to a finite hard wall potential
 \begin{equation}
V=\begin{cases}
    0       & \quad r\leq R_D\\
    V_{0}  & \quad r\geq R_D,\\
  \end{cases}
  \label{eq:Fin_Pot}
\end{equation}
 on the spin-degenerate regimes discussed are shown. Thus, for both the zero field and perpendicular magnetic field regimes, the $\Psi(r=R_D,\phi)=0$ boundary condition is replaced by the continuity condition at the potential interface $\partial_r \ln[\Psi^{r\geq R_D}_{n,l}(r=R_D,\phi)]=\partial_r \ln[\Psi^{r\leq R_D}_{n,l}(r=R_D,\phi)]$\cite{recher2009bound}.
 
\begin{figure}[!t]
    \includegraphics[width=\linewidth]{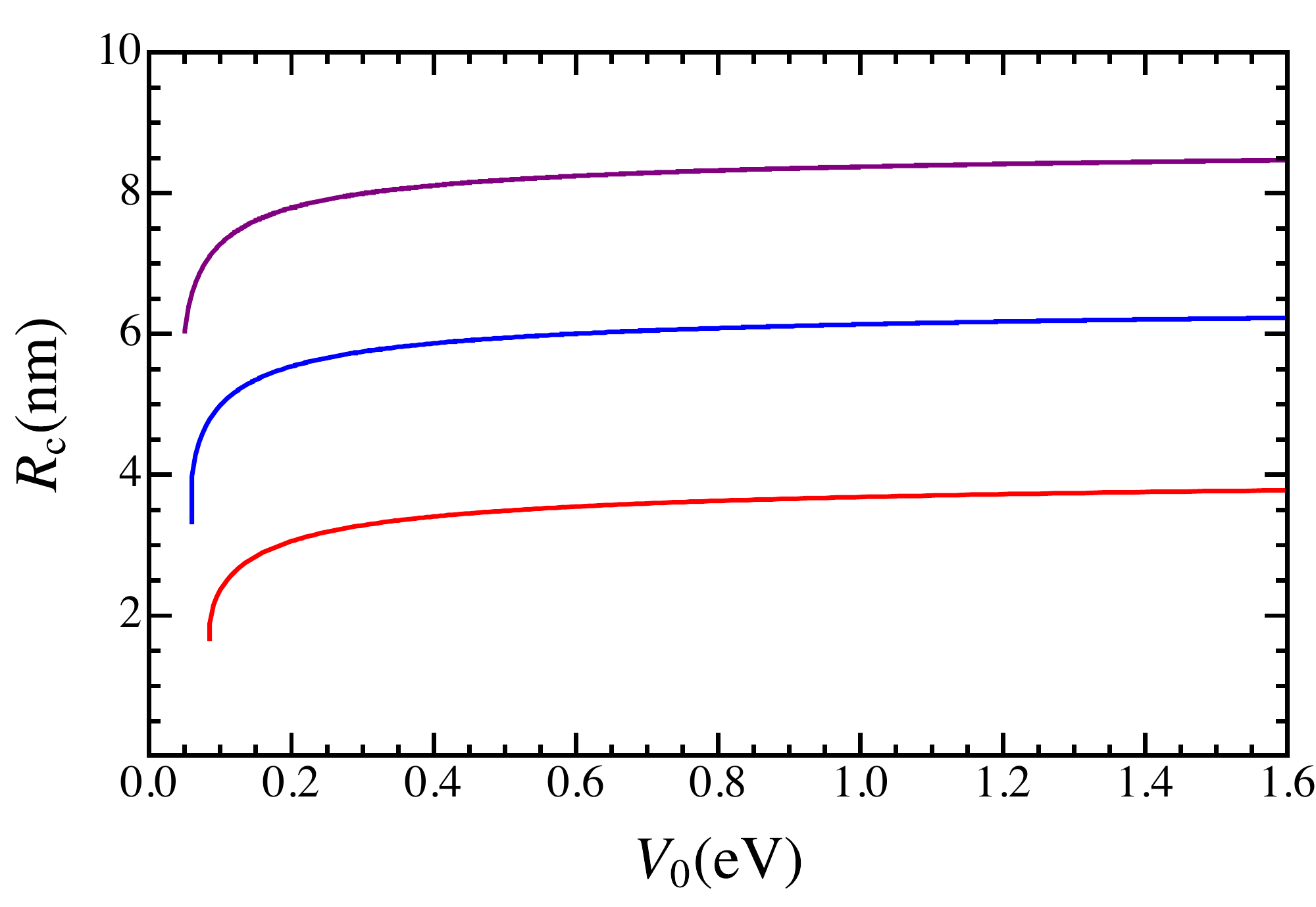}
    \caption{Spin-degenerate critical radii $R_c$ of QD of finite potential height in MoS$_{2}$ monolayers at the ground and first few excited states, red: $n=1$, $l=0$, blue:  $n=1$, $|l|=1$, purple:  $n=1$, $|l|=2$.}
    \label{fig:MoS2_RCrit_VWell_ZF}
\end{figure}

 In zero field the unnormalised radial portions of the wavefunction within and outside of the potential barrier are described as follows
  \begin{equation}
R_{n,l}(r)=\begin{cases}
   J_{|l|} (\epsilon_{n,l}^{\text{in}} r)       & \quad r\leq R_D\\
    e^{\frac{i l \pi}{2}} K_{|l|} (\epsilon_{n,l}^{\text{out}} r)  & \quad r\geq R_D\\
  \end{cases}.
  \label{eq:ZF_Fin_WVFN}
\end{equation}
Here $ K_{l}$ is the $l^{th}$ modified Bessel function of the second kind, $\epsilon_{n,l}^{\text{in}}=\sqrt{2 m_{\text{eff}}^{\tau,s}[E_{n,l}-\tau\Delta_{\text{cb}}s_z]}/\hbar$ and $\epsilon_{n,l}^{\text{out}}=\sqrt{2 m_{\text{eff}}^{\tau,s}[V_{0}-E_{n,l}+\tau\Delta_{\text{cb}}s_z]}/\hbar$. Eigenenergies as a function of potential height may then be numerically calculated by applying the continuity condition to Eq. (\ref{eq:ZF_Fin_WVFN}), 
\begin{equation}
\frac{\epsilon_{n,l}^{\text{in}} J_{|l|+1} (\epsilon_{n,l}^{\text{in}} R_D)}{J_{|l|} (\epsilon_{n,l}^{\text{in}} R_D)}=\frac{\epsilon_{n,l}^{\text{out}} K_{|l|+1} (\epsilon_{n,l}^{\text{out}} R_D)}{K_{|l|} (\epsilon_{n,l}^{\text{out}} R_D)}.
\label{eq:ZF_Fin_Char}
\end{equation}
From this, the fourfold degenerate critical radii as a function of potential height may be calculated, leading to the result shown in Fig. \ref{fig:MoS2_RCrit_VWell_ZF}. The effect of a finite potential is only noticeable at low potential heights $<\unit[100]{meV}$, whereafter a sharp drop in the critical radii is observed.

\begin{figure}[!h]
    \includegraphics[width=\linewidth]{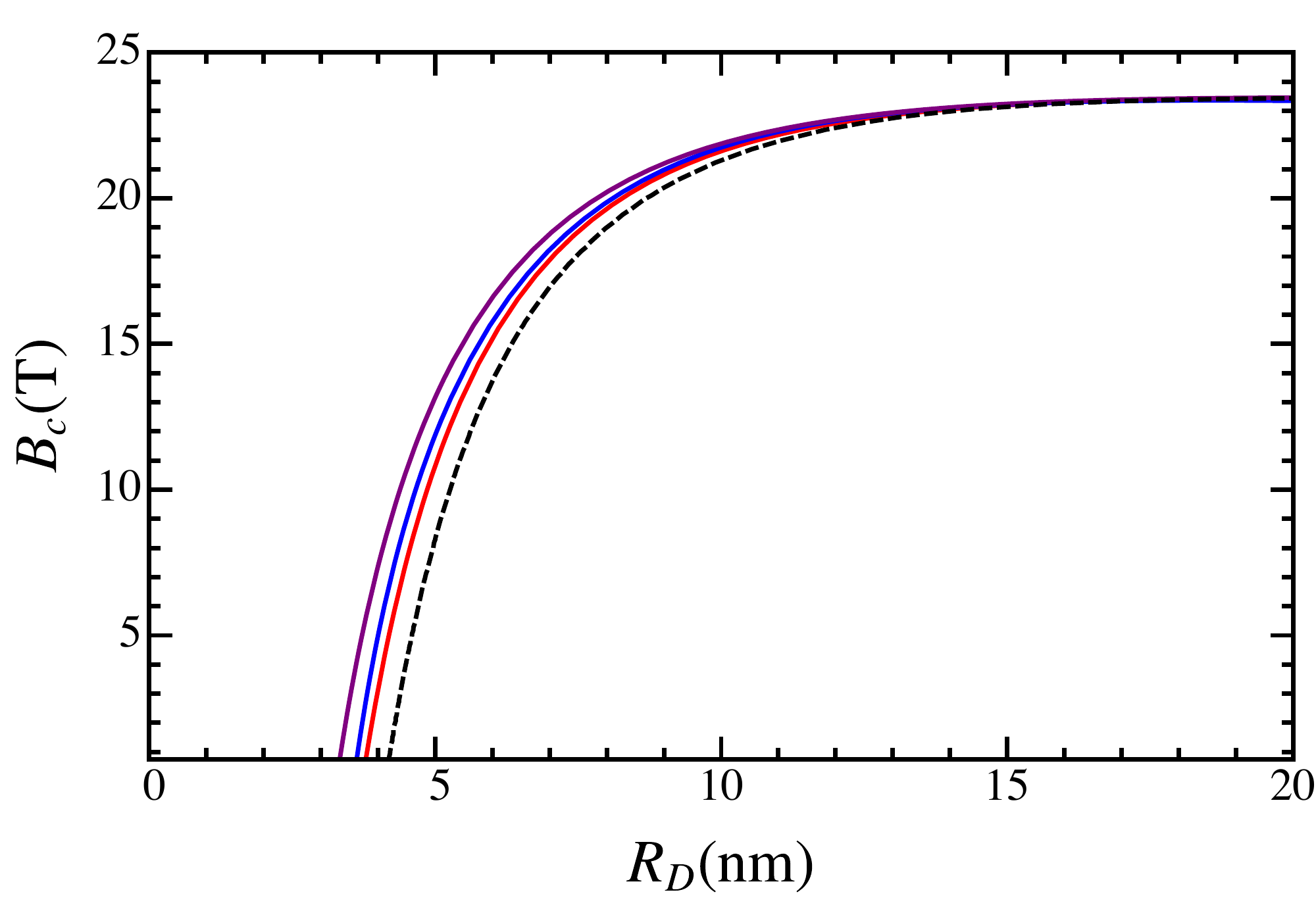}
    \caption{Spin-degenerate critical magnetic field $B_c$ of QD of finite potential heights in MoS$_{2}$ monolayers at the ground of heights $\unit[1]{eV}$ (red), $\unit[0.5]{eV}$ (blue), $\unit[0.25]{eV}$ (purple) and infinite potential (black dashed) for reference}
    \label{fig:MoS2_RCrit_VWell_BPerp}
\end{figure}

Similarly, when a finite potential is considered with an external magnetic field over the QD, the unnormalised radial component of the wavefunction is described as 

\begin{equation}
P_{n,l}(\rho)= \rho^{|l|/2} e^{-\rho/2}\begin{cases}
 M (\tilde{a}^{\text{in}}_{n,l},|l|+1,\rho) & \quad r\leq R_D\\
 U (\tilde{a}^{\text{out}}_{n,l},|l|+1,\rho) & \quad r\geq R_D\\
  \end{cases}
  \label{eq:BPerp_Fin_WVFN}
\end{equation}

\noindent where $U (\tilde{a}^{\text{out}}_{n,l},|l|+1,\rho)$ is Tricomi's hypergeometric function and $\tilde{a}^{\text{in}}_{n,l}$ is the $n^{th}$ numerical solution to the continuity equation at the potential barrier and $\tilde{a}^{\text{out}}_{n,l}=\tilde{a}^{\text{in}}_{n,l}+V_{0}/\hbar \omega_c^{\tau,s}$. The continuity condition may then be applied to achieve the following characteristic equation

\begin{equation}
\begin{split}
&(1+|l|)\tilde{a}_{n,l}^{\text{out}}M(\tilde{a}^{\text{in}}_{n,l},|l|+1,\rho_D)U(1+\tilde{a}_{n,l}^{\text{out}},|l|+2,\rho_D)\\&+\tilde{a}^{\text{in}}_{n,l}M(1+\tilde{a}^{\text{in}}_{n,l},|l|+2,\rho_D)U(\tilde{a}_{n,l}^{\text{out}},|l|+1,\rho_D)=0
\end{split}
\end{equation}
\label{eq:BPerp_Fin_Char}

\noindent from which $\tilde{a}^{in}_{n,l}$ may be numerically extracted and applied to Eq. (\ref{eq:BPerp_total_Energy}) in lieu of $a_{n,l}$. The effect of a finite potential height model on the spin-degenerate regimes of MoS$_{2}$ is shown in Fig. \ref{fig:MoS2_RCrit_VWell_BPerp}. 

A similar effect on the spin degeneracy regimes in shown in both FIGs. \ref{fig:MoS2_RCrit_VWell_ZF} and \ref{fig:MoS2_RCrit_VWell_BPerp}. At shallow potential heights the required critical radius of the dot decreases by $\sim\unit[1-2]{nm}$. However at high magnetic field, there is no discernible difference between the finite and infinite potential solutions. This result will pose little threat to the operation of dots with a single electron charged into the groundstate as the potential height may be selected to be sufficiently high such that little to no difference in the critical radii will be observed. Although, as is discussed in Sec. \ref{sec:QIP}, this effect must be considered when switching to an excited operational electron state by charging.  

\section{Single Quantum Dots as Qubits}
 \label{sec:QIP}

To achieve a pure spin qubit in a single MoS$_{2}$ QD, some considered parameter selection is required to gain a certain robustness of the operational regime. As previously stated in Sec. \ref{sec:PerpB}, a regime with a single electron in the lowest spin-degenerate state either requires a very large external field ($>\unit[20]{T}$) or extreme precision in the QD's radius. This is not ideal, however these problems may be mitigated by charging the dot to operate at higher degenerate states. As can be seen in Fig. \ref{fig:MoS2_Inf_BCrit_RDot_Spin}, at reasonable external fields ($\leq\unit[10]{T}$), for each increasing excited state the necessary QD radius increases  in accordance with Eq. (\ref{eq:ZF_Crit_Rad}). These regimes allowing for larger dot radii are more reliably achieved by gated monolayer QD fabrication methods. Moreover, the $(|l|+l)/2$ term of Eq. (\ref{eq:BPerp_total_Energy}) splits the plateaus of the regime curves shown in Fig. \ref{fig:MoS2_Inf_BCrit_RDot_Spin} into the higher plateaus of the $l\leq0$ and lower $l=1,2,\dots$ plateaus. Therefore, if a charged excited state is chosen as the operational state, the ideal choice would be an $l>0$ angular-state.

Even in the lowest spin-degenerate state, some charging may be required. The operational electron confined to the $K'$ valley is at a higher energy than the two other possible states in the $K$ valley (see Fig. \ref{fig:MoS2_Inf_BPerp_20nm}). Although valley lifetime is expectedly long \cite{yang2015long,sallen2012robust}, eventually the electron will decay out of the higher operational state to these empty states. Also, since each excitation state may be split into four different configurations of spin and valley, the total number of electrons needed to charge the dot up to the desired operational regime is $3+4N$ where $N$ is an integer describing the excitation level of the operational state, i.e. $N=0$ corresponds to the groundstate $n=1$ $l=0$, $N=1$ corresponds to the first excited state $n=1$ $l=-1$ etc. The direct band gap of monolayer MoS$_{2}$ is $\sim\unit[1.8]{eV}$ \cite{mak2010atomically}, and current advances in gated QD nanostructures in MoS$_{2}$ give a charging energy of $\unit[2]{meV}$ at a dot radius of $\unit[70]{nm}$ \cite{wang2016engineering}. This result was said to align well with the self capacitance model \cite{wang2016engineering,kouwenhoven2001few,hanson2007spins}, therefore, using said model, the charging energy at desired radii for spin-degenerate regimes ($\sim\unit[10]{nm}$) may be approximately shown to increase to $\sim\unit[14]{meV}$. This is however a broad approximation, therefore further study of the perturbation of the energy levels due to Coulomb interaction mediated by the Keldysh potential\cite{chernikov2014exciton} is warranted, however such effects are spin and valley independant and should only serve as a renormalisation of the effects studied here. These considerations do however limit the choice of excited operational states, as is evident in Fig. \ref{fig:MoS2_RCrit_VWell_ZF}, at highly charged states relative to the potential height and band gap, the critical radii will be compromised.

Additionally, ferromagnetic substrates may be employed to enhance the valley splitting due to an external magnetic field. Recent experiments demonstrate an effective $\sim\unit[2]{T}$ addition to the magnetic field inducing valley Zeeman splitting in WSe$_{2}$ monolayers on EuS ferromagnetic substrate \cite{zhao2016enhanced}. Such techniques may be employed to reduce the necessary external field strength to reasonable quantities.

An alternative quantum confinement method with TMDC monolayers has been proposed by way of heterostructures consisting of islands of one form of Mo based TMDC within a sea of a the corresponding W based monolayer \cite{liu2014intervalley,PhysRevB.93.045313}, or by sufficiently small free standing flakes\cite{pavlovic2015electronic}. While such methods offer quantum confinement on the desired scale, high inter-vally coupling terms are introduced at small dot radii due to edge effects, offering a decoherence channel to the system. Additionally, such structures offer scalability challenges such as the lack of a method of adjusting the exchange coupling if the proposed model is extended to a double QD system. However, such studies of quantum confinement in TMDCs pay close attention to the effect of dot shape, a consideration omitted here for simple symmetry considerations, but could yet warrant consideration in further research. 

With a suitable operational regime selected, operation of the spin qubit is relatively straightforward. The energy gap between the up and down spin computational basis is tuneable by the external magnetic field, while Bychkov-Rashba spin orbit coupling induced by an external electric field perpendicular to the device may be used to provide off diagonal spin coupling terms in the spin Hilbert space \cite{PhysRevX.4.011034}. 

\section{Summary}
\label{sec:Summary}

Overall, given selection of a proper operational regime and reasonable accuracy in QD fabrication at low radii, MoS$_{2}$ monolayer QDs do offer novel pure spin qubits in 2D semiconductors. Overcoming the Kramers pairs of gated QDs on TMDC monolayers is explored, as to achieve operational regimes of pure spin qubits, thus avoiding the problem of achieving valley state mixing and  low valley coherence times. Zero field fourfold spin-valley degeneracy was demonstrated to be achievable in Mo based TMDC monolayers, unlike their W based counterparts, at low QD radii whilst spin degeneracy solely within a given valley was shown be achieved by application of a sufficiently high external magnetic field perpendicular to the dot. Regime restrictions for spin-degenerate MoS$_{2}$ QDs have been shown, demonstrating radially sensitive low external field regimes which may be made to be more robust when charged into higher operational states and enhanced valley-Zeeman splitting substrates. Switching from an infinite to a finite potential barrier model did demonstrate a drop in the expected values of spin-degenerate critical radii, but only at particularly low barrier heights. In addition to the moderate expected charging energy this somewhat limits the usefulness of highly charged operational states, but will not substantially effect operation at the first few excited states. To conclude, a theoretical demonstration of QD radius dependant spin-orbit effects in TMDC monolayers is given along with descriptions of possible methods of implementing novel pure spin qubits on two-dimensional semiconductor crystals.

\section{Acknowledgements}
\label{ref:Acknowledgements}

We acknowledge helpful discussions with A. Korm\'{a}nyos, A. Pearce, M. Ran\v{c}i\'{c} and M. Russ and funding through both the European Union by way of the Marie Curie ITN Spin-Nano and the DFG through SFB 767.

\bibliography{bibliography}

\end{document}